\documentclass[pdflatex,sn-mathphys-num,iicol]{sn-jnl}%

\usepackage{graphicx}%
\usepackage{multirow}%
\usepackage{amsmath,amssymb,amsfonts}%
\usepackage{amsthm}%
\usepackage{mathrsfs}%
\usepackage[title]{appendix}%
\usepackage{xcolor}%
\usepackage{textcomp}%
\usepackage{manyfoot}%
\usepackage{booktabs}%
\usepackage{array}
\usepackage{algorithm}%
\usepackage{algorithmicx}%
\usepackage{algpseudocode}%
\usepackage{listings}%
\usepackage{makecell}
\usepackage{threeparttable}
\usepackage{multirow}
\usepackage{graphicx}
\usepackage{booktabs}
\usepackage{tabularx}
\usepackage{siunitx} 

\theoremstyle{thmstyleone}%
\theoremstyle{thmstyletwo}%

\theoremstyle{thmstylethree}%

\begin{document}

\title[Article Title]{Towards user-centered interactive medical image segmentation in VR with an assistive AI agent}


\author*[1]{\fnm{Pascal} \sur{Spiegler}}\email{pascal.spiegler@mail.concordia.ca}

\author[1]{\fnm{Arash} \sur{Harirpoush}}

\author*[1]{\fnm{Yiming} \sur{Xiao}}\email{yiming.xiao@concordia.ca}

\affil[1]{\orgdiv{Department of Computer Science and Software Engineering}, \orgname{Concordia University}, \orgaddress{\city{Montreal}, \state{Quebec}, \country{Canada}}}


\abstract{Crucial in disease analysis and surgical planning, manual segmentation of volumetric medical scans (e.g. MRI, CT) is laborious, error-prone, and challenging to master, while fully automatic algorithms can benefit from user feedback. Therefore, with the complementary power of the latest radiological AI foundation models and virtual reality (VR)'s intuitive data interaction, we propose SAMIRA, a novel conversational AI agent for medical VR that assists users with localizing, segmenting, and visualizing 3D medical concepts. Through speech-based interaction, the agent helps users understand radiological features, locate clinical targets, and generate segmentation masks that can be refined with just a few point prompts. The system also supports true-to-scale 3D visualization of segmented pathology to enhance patient-specific anatomical understanding. Furthermore, to determine the optimal interaction paradigm under near-far attention-switching for refining segmentation masks in an immersive, human-in-the-loop workflow, we compare VR controller pointing, head pointing, and eye tracking as input modes. With a user study, evaluations demonstrated a high usability score (SUS=90.0±9.0), low overall task load, as well as strong support for the proposed VR system's guidance, training potential, and integration of AI in radiological segmentation tasks.}




\maketitle

\section{Introduction}\label{sec1}

Medical image segmentation is a critical task in clinical diagnosis and treatment planning, particularly for identifying and quantifying abnormalities such as tumors, stroke lesions, and other pathological anomalies. The process involves producing segmentation masks that delineate and highlight regions of interest to provide a basis for further analysis, treatment decisions, and longitudinal tracking. However, traditional workflows for medical image segmentation are time-consuming and labor-intensive, typically requiring experts to manually annotate up to hundreds of 2D slices to isolate structures within 3D MRI or CT scans. Furthermore, while segmenting pathologies (e.g., tumor) accurately is itself a demanding skill, it often requires extensive hours of supervision and training to develop diagnostic confidence and anatomical precision \cite{bruno2015radiology}. Finally, visualizations of these annotations are similarly challenging: clinicians either scroll through superimposed binary masks on 2D slices across the axial, sagittal, and coronal planes, or view 3D reconstructions rendered on flat screens, both of which lack real spatial context and a true sense of scale.

While virtual reality (VR) can offer more intuitive 3D medical data visualization and interaction, especially under high spatial constraints (e.g., in the clinic), recent developments in foundation artificial intelligence (AI) models, such as vision-language models (VLMs) have demonstrated early promise to further enhance the efficiency, accuracy, and interactability for tasks in VR in the form of AI agents \cite{konenkov2024vr, behravan2025generative}. As an alternative to conventional brush painting-based segmentation paradigms, imagine a workflow that unites the efficiency of AI with the spatial interaction advantages of a virtual environment: a user reviews a brain tumor MRI in VR assisted by an AI agent. First, the agent guides them towards a representative tumor slice. When the user confirms this slice contains a tumor, they issue a simple voice command, triggering the AI agent to highlight the tumor and provide a detailed patient-specific radiological description. Crucially, this interaction goes beyond the role of a typical conversational assistant: rather than passively responding to dialogue, the AI agent actively executes a sequence of actions, from volumetric segmentation to case-based guidance and full 3D rendering with minimal input. If necessary, the user can easily refine the segmentation using natural inputs, including head pointing, gaze, or handheld controllers, and finally, the corrected mask is rendered in 3D at true scale, offering enhanced spatial interpretation of the tumor’s dimensions. Notably, as current deep learning algorithms for radiological segmentation still require human quality assurance \cite{budd2021survey, vasquez2024human, jin2024quality}, such a human-in-the-loop approach preserves user oversight while dramatically improving speed, accuracy, and spatial understanding through immersive AI assistance. 

In our work, we present such a VR-based, AI-assisted medical image segmentation system that supports medical image review for both diagnostic decision-making and education. To the best of our knowledge, this system is the first of its kind, not only automating segmentation across image slices via a conversational AI agent, but also investigating optimal interaction paradigms for human-in-the-loop mask refinement. Our novel contributions are as follows:
\textbf{First}, we systematically investigate optimal interaction paradigms for human-in-the-loop segmentation refinement that transitions between proximal corrections and distal menu interactions, comparing natural inputs from handheld controllers, head pointing, and gaze. \textbf{Second}, we introduce a new 3D segmentation algorithm based on the BiomedParse and SAM2 foundation models which reduces mask drift from noise during slice-to-slice annotation propagation in medical images. \textbf{Finally}, we present SAMIRA, a conversational AI agent that assists with segmenting 3D radiological scans via voice commands, provides interactive guidance, supports iterative refinement to preserve expert oversight, and enables life-scale 3D visualization of results in immersive VR.

\section{Related Work}\label{sec2}

Although research on AI assistants for clinical training in VR is still in its infancy, early work has begun to demonstrate its potential to support clinical workflows and healthcare education. For example, Liaw et al.~\cite{liaw2023ai} developed a conversational AI assistant within a VR simulation for sepsis training, which matched human-controlled scenarios in clinical and communication performance and led to significantly higher test scores. Furthermore, Chheang et al.~\cite{chheang2024anatomy} introduced a generative AI-based virtual assistant in a VR anatomy education environment, highlighting the potential of such assistants to support anatomy education. Although these systems leverage large language models (LLMs) to provide educational support via dialogue, they primarily function as virtual assistants focused on delivering information. In contrast, AI agents should autonomously execute sequential tasks in response to high-level user prompts, for example, as in our scenario, performing segmentation, providing clinical context via text and images, and rendering 3D models. Retrieval-Augmented Generation (RAG) often powers such agents by retrieving relevant examples from a structured knowledge database and combining similarity search results with generative models to offer refined responses to queries. This contextualized guidance is particularly useful for medical tasks, where visual patterns vary across cases. To our knowledge, no prior work has integrated AI agents into immersive VR for interactive 3D radiological segmentation.

Recent development in foundation models has paved the way for interactive medical image analysis with multi-modal inputs. One such model, BiomedParse\cite{zhao2025biomedparse}, is a Transformer-based vision-language model that uses separate encoders for medical images and clinical text data. Trained on 1.2 million paired 2D radiological images and reports, it supports detection, classification, and segmentation of 82 clinical concepts (e.g. tumor) across 9 imaging modalities using natural language prompts. On the other hand, SAM2\cite{kirillov2024sam2} is a foundation model for segmenting objects in images and video using user-provided point and/or box prompts that enables prompt-based segmentation mask refinement and leverages a memory encoder to propagate masks across video frames. Although developed for natural image domains, SAM2's ability to handle sequential image data positions itself as a promising tool for 3D medical imaging segmentation, where volumetric scans, such as CT or MRI can be viewed as series of 2D images analogous to videos \cite{shen2025interactive3dmedicalimage, zhu2024medical}. This opens up the possibility of combining BiomedParse’s language-driven label mask generation with SAM2’s interactive label refinement and propagation capabilities to enable human-in-the-loop segmentation workflows for 3D medical imaging, an underexplored area, particularly in immersive VR environments. While prior VR segmentation systems have allowed users to paint regions of interest using hand gestures~\cite{gao2022vr, yonker20193d} or have used predefined anatomy-specific deep learning models~\cite{gonzalez2020nextmed}, none have integrated foundation segmentation models with agent-driven guidance to support interactive refinement.

In immersive medical image segmentation with SAM2, users will need to frequently switch their visual and motor attention between the image being annotated and a spatially decoupled user interface (UI) menu for triggering actions (e.g., toggling between negative/positive prompts, resetting prompts, etc.), thus creating a dual‑focus challenge. Previously, Rashid et al. \cite{rashid2011proximal} compared proximal (on-device) versus distal (remote) widget selection in distributed user interfaces and found that proximal methods were faster and preferred for complex, multi‑step tasks, while distal methods yielded lower error rates for simpler interactions, suggesting that high‑precision tasks like segmentation prompt placement may benefit from proximal selections and that simpler menu selections may benefit from being performed at a distance. In our case, the interactive refinement of AI-predicted segmentations with SAM2 in VR necessitates identifying optimal interaction paradigms for completing the task under attention-switching. \textit{However, the ideal interaction paradigm (e.g., controller, head pointing, or eye-tracking) for attention switching between proximal and distal displays remains unexplored.} Sidenmark et al. \cite{sidenmark2023gaze} conducted a head‑mounted VR study comparing gaze, head, and controller pointing for dynamically revealed target selection, showing that both gaze‑ and controller‑based pointing significantly outperformed head pointing in terms of speed and precision, though they did not explore static dual‑panel switching scenarios. Luro et al. \cite{luro2019comparative} compared eye tracking with hand‑controller aiming tasks in VR and reported that controllers achieved the highest placement accuracy, while gaze‑based selection felt more natural, but suffered from increased selection accuracy variability. Studies of multi‑depth targeting, such as the experiment of Schultheis et al. \cite{fernandes2024looking} revealed that eye‑based selection can achieve higher throughput across varying depth planes, but controller input offers more stable performance when precision is critical. Furthermore, Xu et al.'s evaluation of text‑selection techniques in VR \cite{xu2022evaluation} demonstrated that head‑pointing with click confirmation strikes a balance between speed and accuracy, ranking just behind controller pointing in speed while maintaining a minimal task load, positioning it as a possibly viable paradigm for attention switching interactions. With these previous insights, we will investigate the optimal interaction paradigms for point placement and menu selection under attention switching conditions for our application.

\section{System Overview}\label{sec3}

\begin{figure} [h!]
    \centering
    \includegraphics[width=1\linewidth]{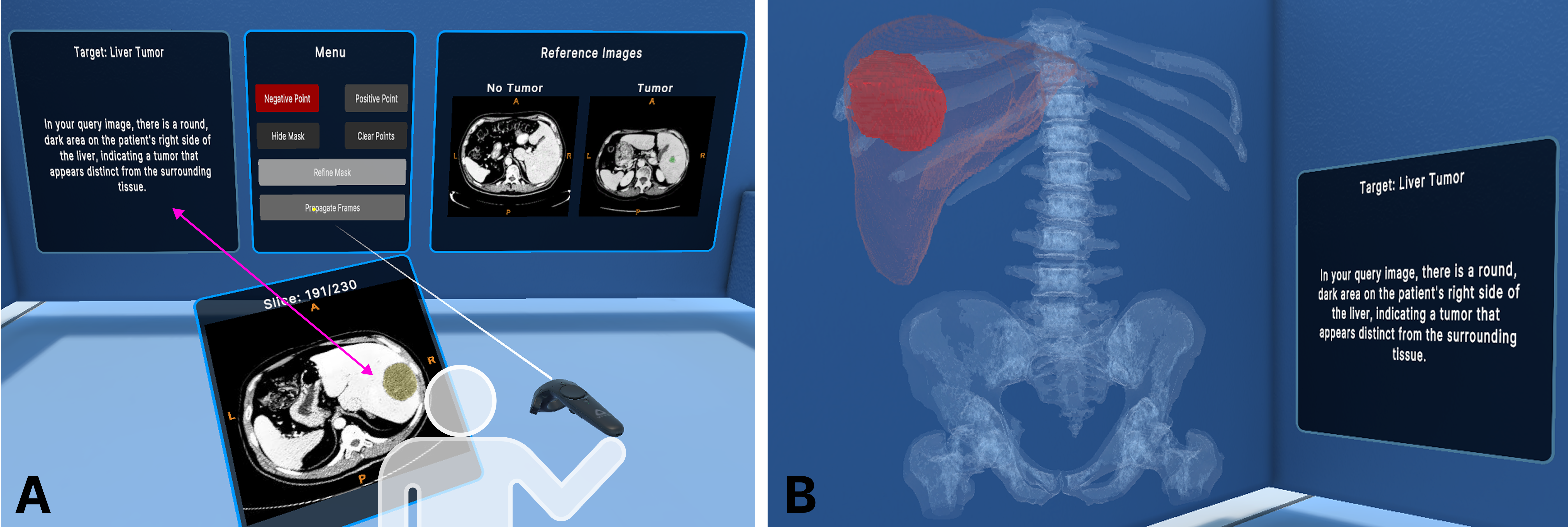}
    \caption{A. The AI agent generates an initial segmentation of a liver tumor in CT and provides guidance using reference images and patient-specific pathology explanations. B. The final refined 3D visualization is rendered as a large, spherical, high-contrast liver tumor in red, overlaid on anatomical structures, to scale. Few refinements are expected due to the simple shape.}
    \label{fig:SAMIRA_Overview}
\end{figure}

To address the aforementioned issues, we present \emph{SAMIRA} (Segmentation Assistant for Multimodal Interaction and Radiological Analysis), a novel conversational agent designed to support human-in-the-loop 3D medical image segmentation by generating segmentation masks, enabling efficient mask refinement, and providing radiological guidance through speech and reference images. Figure~\ref{fig:SAMIRA_Overview} illustrates the system’s key components during the segmentation of a liver tumor in a CT scan.

All VR development was conducted in Unity 2022.3.f1 on a desktop equipped with an NVIDIA RTX 3090 and an HTC Vive Pro Eye VR headset. In the VR system, medical image slices are displayed on a virtual panel anchored to a VR controller in the user’s hand, allowing flexible repositioning for detailed inspection. Users navigate through image slices by rotating the thumbpad on the HTC Vive controller, clockwise to advance and counter-clockwise to reverse. Each 60° rotation corresponds to a single slice, reinforced by a haptic pulse that provides tactile feedback at each transition.

The menu interface is structured into three panels (see Figure~\ref{fig:SAMIRA_Overview}): on the left panel, AI-driven guidance is provided as text and synthesized speech for the given segmentation task. In the middle panel, interactive controls with functional buttons allow users to issue voice commands, refine predicted masks, propagate segmentations across slices, and render results in 3D. On the right panel, reference images are retrieved in real time based on the user’s currently viewed slice based on RAG from an existing knowledge repository. On this panel, the system presents anatomically similar image examples with and without the target structure (e.g., tumor), helping users locate pathology and build a clearer understanding of its visual characteristics.

Using the AI agent’s guidance and slice-scrolling mechanism, users can find the target pathology, initiate segmentation via voice commands, refine masks using efficient natural inputs, then render a final segmentation into a true-to-scale 3D visualization to gain spatial understanding of pathologies.

All real-time inference (e.g., RAG retrieval), segmentation generation, point-prompt-based segmentation refinement, mask propagation, and 3D mesh creation are realized by SAMIRA and coordinated via two dedicated Python WebSocket servers: a local rendering server and a dedicated inference server. This is to ensure responsive, low-latency AI assistance and integration between Unity and the supporting AI models. However, future similar systems can benefit from cloud-based setups.

\vspace{0.1cm}
\noindent
\textbf{Local Rendering Server:} The server ran on the same desktop as the Unity client. It receives binary segmentation masks, scales the lesions using medical image metadata, and returns a \texttt{.obj} mesh for VR visualization.

\vspace{0.1cm}
\noindent
\textbf{Dedicated Inference Server:} A separate Ubuntu 22.04.2 LTS server was set up with a dedicated NVIDIA RTX 3090 to host SAMIRA's RAG and foundational segmentation models (BiomedParse, SAM2). It handles voice commands, reference retrieval, textual guidance, text-based segmentation generation, mask refinement, and mask propagation across frames (with SAM2).

\section{Methods and Materials}\label{sec4}

\subsection{Interaction Paradigm Evaluation}
\begin{figure*}[!htbp]
    \centering
    \includegraphics[width=\textwidth]{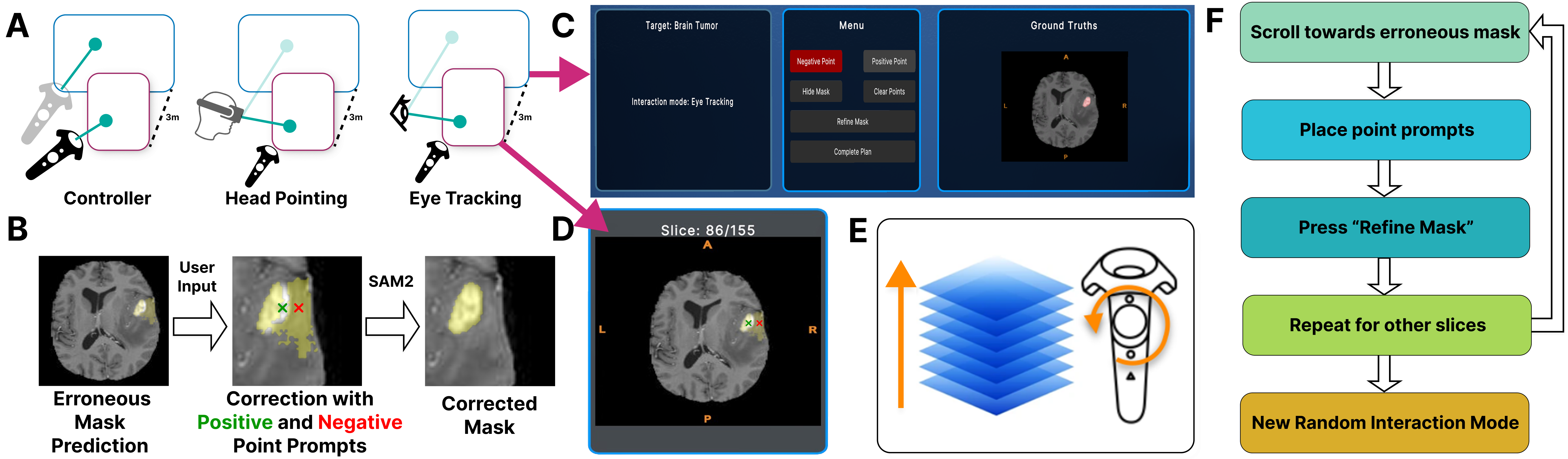}
    \caption{System setup for user interaction paradigm evaluation under attention switching. A. Three interaction paradigms: controller ray, head pointing, and eye tracking.
B. Users correct erroneous masks using positive (\textcolor{teal}{green}) and negative (\textcolor{red}{red}) point prompts, refined by SAM2.
C. In-VR interface for prompt selection and real-time ground truth reference.
D. Medical image display with current slice and segmentation overlay.
E. Controller-based slice scrolling.
F. Interaction paradigm evaluation segmentation workflow.}
    \label{fig:Interactions_Overview}
\end{figure*}

Before evaluating the full VR system with our AI agent, SAMIRA, we aimed to first reveal the optimal interaction paradigm for point prompt placement and menu selection under attention switching between the proximal handheld image display and the distal menu interface. To accomplish this, we designed an experiment to perform SAM2-based segmentation mask refinement using three distinct paradigms: \emph{Controller}, \emph{Head Pointing}, and \emph{Eye Tracking}. The detailed setup is shown in Fig. \ref{fig:Interactions_Overview}. As a potential middle ground between \textit{Controller} and \textit{Eye tracking}, \textit{we hypothesize that \textit{Head Pointing} is the optimal interaction paradigm for our intended application, with the consideration of task load, accuracy, and efficiency}. 

\vspace{0.2cm}
\noindent
\textbf{Controller} pointing is a staple method for object and menu selection in many VR applications. We employed standard controller-based ray-casting, where a visible ray extends from the tip of the controller to intercept with the image under analysis and the menu for point prompt annotation and button clicking, respectively. Here, confirmation of selection is achieved by pressing the trigger button of the controller.

\vspace{0.2cm}
\noindent
\textbf{Head Pointing} adopts an invisible ray forward from the center of the user’s headset, aligning with the head's orientation. The interception of the ray and the image/menu is represented by a visible dot, and the controller trigger button is used for placement/selection confirmation.  

\vspace{0.2cm}
\noindent
\textbf{Eye Tracking} utilizes the integrated eye-tracking hardware of the HTC Vive Pro Eye headset to cast an invisible ray based on the user's gaze direction. Similar to head pointing, a visible dot is used to indicate the gaze point on the image under analysis and the menu. Based on previous studies \cite{hellum2023assessment, mutasim2021pinch}, we continue to use the trigger button to confirm selection. To mitigate the adverse impacts of natural eye jitter on selection precision and user experience, we apply exponential smoothing to the normalized 3D gaze vectors during every frame using a smoothing factor of \( \alpha = 0.2 \). The gaze ray is computed as follows:

    \begin{equation}
        \mathbf{r}_{\text{smooth}} = (1 - \alpha) \cdot \mathbf{r}_{\text{previous}} + \alpha \cdot \mathbf{r}_{\text{current}}
        \label{eq:smoothing}
    \end{equation}

    \noindent
    where \( \alpha = 0.2 \) is the smoothing factor, \( \mathbf{r}_{\text{previous}} \) is the smoothed direction from the previous frame, and \( \mathbf{r}_{\text{current}} \) is the normalized average of the left and right eye gaze direction vectors, originating at the midpoint of the two eyes.

\vspace{0.2cm}        
The overall interaction paradigm evaluation workflow is summarized in Figure~\ref{fig:Interactions_Overview}F, where the user begins at the middle slice of a segmented 3D scan and scrolls through the volume to identify and correct segmentation errors. Three fixed large panels (Figure~\ref{fig:Interactions_Overview}C) are positioned three meters in front of the user, different from the panel displays for the full interactive segmentation workflow (Figure~\ref{fig:SAMIRA_Overview}): the left panel displays the current interaction mode (controller, head pointing, or eye tracking), the central panel provides interactive controls (i.e., functional buttons), and the right panel shows the ground truth segmentation (red, 40\% opacity overlaid on the image) for the current slice under study. Here, the ground truth segmentation is used to define a consistent reference for refinement and objective evaluation of segmentation accuracy across participants in the user study. Users were instructed to correct the provided masks until they visually matched the ground truths. The central panel includes six buttons. ``Positive Point" and ``Negative Point" buttons allow users to place point prompts that add missing regions or remove excess segmentation, respectively. The ``Hide Mask" button toggles the visibility of both the segmentation under work and ground truth to better assess tissue boundaries. The ``Clear Points" button erases all placed point prompts without modifying the current mask. The ``Refine Mask" button sends the current slice and point prompts to the inference server to update the segmentation. Finally, ``Complete Plan" finalizes segmentation refinement and advances to the next interaction paradigm (\textit{Controller}, \textit{Head Pointing}, or \textit{Eye Tracking}), selected randomly, allowing within-participant comparisons of performance across paradigms.

To evaluate interaction paradigms under attention-switching for AI-facilitated segmentation refinement, we used a T1c MRI scan from the publicly available BRATS brain tumor dataset~\cite{menze2014multimodal}, which was intensity-normalized and converted into a sequence of 155 axial JPEG image slices. All slices along with 44 slice-wise ground truth masks and 16 intentionally corrupted slice-wise masks were included. The corrupted masks represented common failure cases, such as incomplete tumor coverage, over-segmentation, missing regions, or false positives. The quality of the initial segmentation, measured using the 3D Dice score, was assessed at 0.91. 3D Dice is a standard metric for volumetric medical segmentation that quantifies the spatial overlap between the predicted mask $A$ and the ground truth $B$ as:
    \begin{equation}
    \text{Dice}(A, B) = \frac{2|A \cap B|}{|A| + |B|}
    \label{eq:dice}
    \end{equation}
A Dice score of 1 indicates perfect agreement; 0 indicates no overlap.

\subsection{SAMIRA - User Interface and Workflow}
The full workflow of our interactive medical image segmentation leverages SAMIRA, our assistive AI agent via the VR user interface described in Section~\ref{sec3}. The interface enables SAMIRA to operate on unseen cases without ground truth segmentations, and supplies reference images and case-relevant descriptive guidance in real time. This supports the potential for future clinical and educational applications beyond the controlled evaluation settings showcased in our study.

\begin{figure*}[!htbp]
    \centering
    \includegraphics[width=\textwidth]{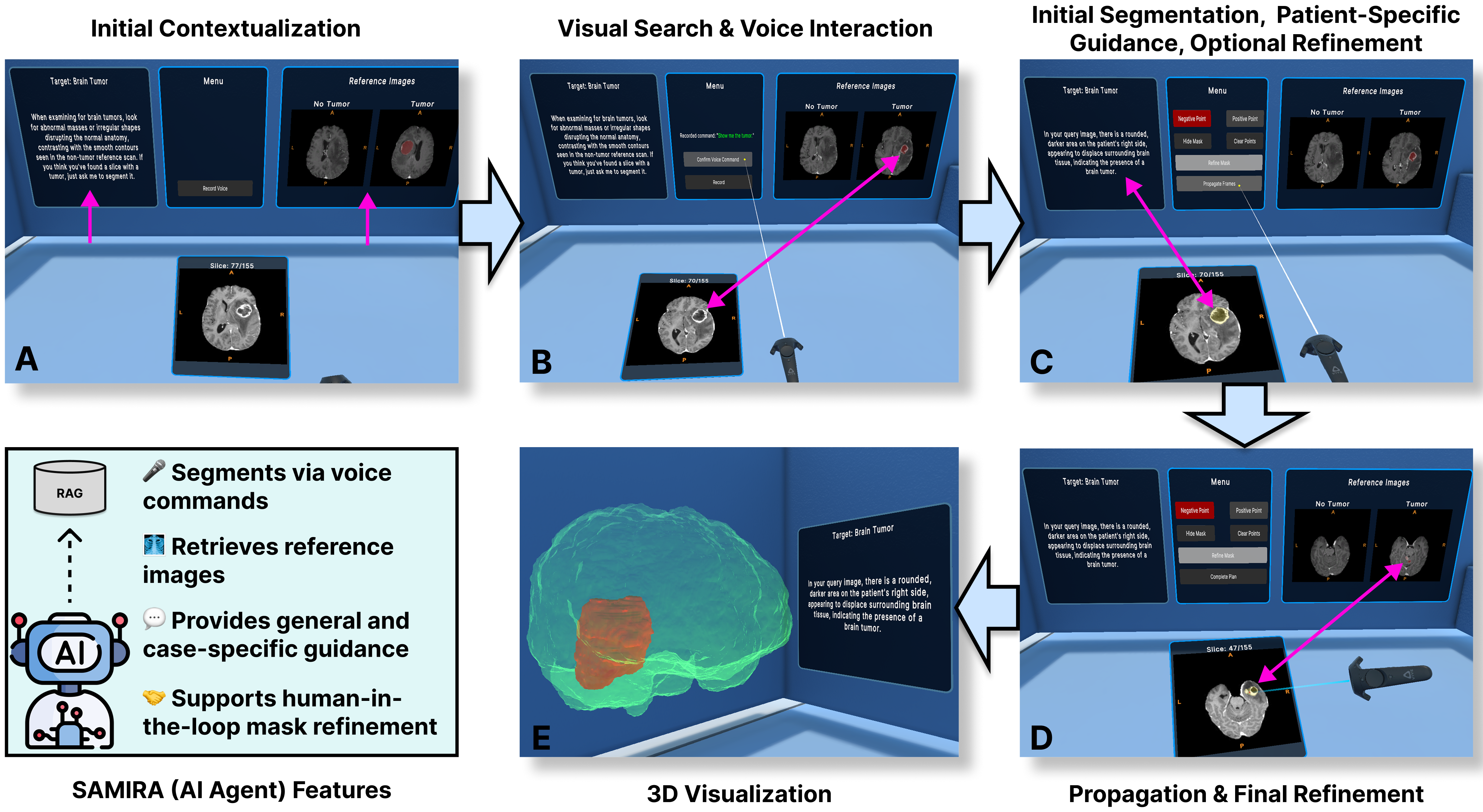}
    \caption{Demonstration of workflow for the proposed AI-assisted interactive medical image segmentation in VR. Users begin by reviewing AI-generated textual guidance and visually similar reference slices (A), then navigate the volume to find the tumor. Once found, they issue a voice command to segment the tumor (B). Next, the agent predicts a mask and a patient-specific description of the tumor (C). If necessary, users can edit this mask, then propagate it across frames. Finally, users review all predicted masks and place point prompts to refine the masks (D). Upon completion, the final segmented structure is rendered in true 3D scale over the patient's anatomy (E).}
    \label{fig:FW_Overview}
\end{figure*}

For demonstration and evaluation purposes, the segmentation workflow was applied to a brain tumor MRI scan from another dataset, Pretreat-MetsToBrain-Masks ~\cite{ramakrishnan2024large}, and a liver tumor CT scan from the LiTS dataset~\cite{bilic2023liver}. These two datasets were deliberately chosen to assess generalization beyond the training distribution of the underlying models, since the model we are introducing, BiomedParse was trained on the BRATS dataset used in the  interaction paradigm test. While the examples focused on tumors, the workflow is compatible with any of the 82 clinical targets supported by BiomedParse. The brain tumor example was selected as the more difficult case, with branching regions (see Figure \ref{fig:FW_Overview}E), whereas the liver tumor case was expected to be easier to segment, with a smoother, spherical structure in which less refinement is expected (see Figure \ref{fig:SAMIRA_Overview}).

 The full workflow is illustrated in Figure \ref{fig:FW_Overview} and consists of the following sequential steps:

\vspace{0.1cm}
\noindent
\textbf{Step 1. Initial Contextualization:} A handheld image display begins at the middle slice of the brain or the liver scan. The AI agent introduces general guidance for segmenting the clinical target on the left panel (Figure \ref{fig:FW_Overview}A) using a RAG pipeline. To ground users’ visual understanding, the system retrieves and displays a positive example (with pathology) and a negative example (without pathology) on the right panel, allowing for contrastive comparison. Both examples bear anatomical similarity to the image slice under study. Then, general pathology explanations are delivered via Google’s Text-To-Speech API in a female voice, and are displayed as text on the left panel. At this stage, the middle panel has one button that says ``Record Voice".


\vspace{0.1cm}
\noindent
\textbf{Step 2. Visual Search:} Users scroll through the image slices in search of the target region and the agent continuously updates the right panel reference panel with visually similar tumor and non-tumor cases retrieved from other patients. The contextual guidance in Step 1 helps less experienced users with deciding whether the current slice contains the pathology of interest (see Figure \ref{fig:FW_Overview}B). 

\vspace{0.1cm}
\noindent
\textbf{Step 3. Voice Interaction:} Users initiate segmentation by clicking the ``Record Voice" button and issuing natural language requests to the AI agent. Spoken input is transcribed in real time using the Microsoft Azure Speech SDK, with recognized commands displayed on the middle panel (Figure \ref{fig:FW_Overview}B). After voice recognition, users can press the ``Record Voice" button again to re-record, or press ``Confirm Voice Command" to pass their command to the inference server.

\vspace{0.1cm}
\noindent
\textbf{Step 4. Initial Segmentation and Patient-Specific Guidance:} The AI agent returns an initial segmentation mask at the selected 2D slice and uses its RAG pipeline to provide additional spoken, case-specific diagnostic context (Figure \ref{fig:FW_Overview}C). The middle panel updates with the same refinement options as the \textit{Interaction Paradigm Evaluation} interface and an additional button ``Propagate Frames".

\vspace{0.1cm}
\noindent
\textbf{Step 5. Refinement and Propagation:} Image slice scrolling is disabled at this stage, forcing the users to focus on the selected 2D slice of interest to achieve the best segmentation accuracy via point-prompt-based refinement if necessary. This is because it will be used to seed the automatic SAM2-based label propagation to obtain the full 3D segmentation. Once satisfied with the current slice, the users initiate mask propagation across slices by pressing the ``Propagate Frames" button. As the inference server returns produced segmentation masks in sequence, the image display automatically updates the slices in the order that the masks are produced in real-time, enabling users to visually assess the segmentation results and quickly judge whether additional refinements will be needed.

\vspace{0.1cm}
\noindent
\textbf{Step 6. Final Refinement of Predicted Masks:} Users review the propagated masks across slices, applying additional point-prompt-based refinements wherever segmentation errors remain (Figure \ref{fig:FW_Overview}D). During this stage, the ``Propagate Frames" button is replaced by a ``Complete Plan" button. When pressed, a confirmation screen appears to prevent accidental submission, ensuring that users have fully completed their corrections before finalizing the segmentation plan. Upon completion, the final segmentation masks are saved.

\vspace{0.1cm}
\noindent
\textbf{Step 7. 3D Visualization:} Upon task completion, a request is sent to the local Python-based websocket server with the path to the saved segmentation result. The server uses the \textit{Marching Cubes} algorithm to extract a polygonal mesh from the 3D segmentation and generates a corresponding \texttt{.obj} file representing the tumor pathology. To ensure anatomically accurate scaling, the voxel resolutions are extracted from the medical image's metadata and applied during mesh generation. The resulting mesh is then loaded back into Unity at runtime and rendered at true-to-life scale (Figure \ref{fig:FW_Overview}E). For brain tumor visualization, a threshold of 500 is applied to extract the brain surface from the scan, while liver tumor visualization uses a threshold of 150 to capture relevant anatomical structures such as the spine and ribcage. Additionally, liver tissue from the same patient was rendered using data from the LiTS dataset. In the future, it can easily be generated with liver segmentation models, or even interactively with our workflow.

\subsection{SAMIRA - Segmentation and RAG}
\noindent
SAMIRA leverages AI foundation models and Retrieval-Augmented Generation to assist the designated interactive medical image segmentation task.

\subsubsection{\textbf{Interactive segmentation with foundation models}}

\begin{figure} [h!]
    \centering
    \includegraphics[width=1\linewidth]{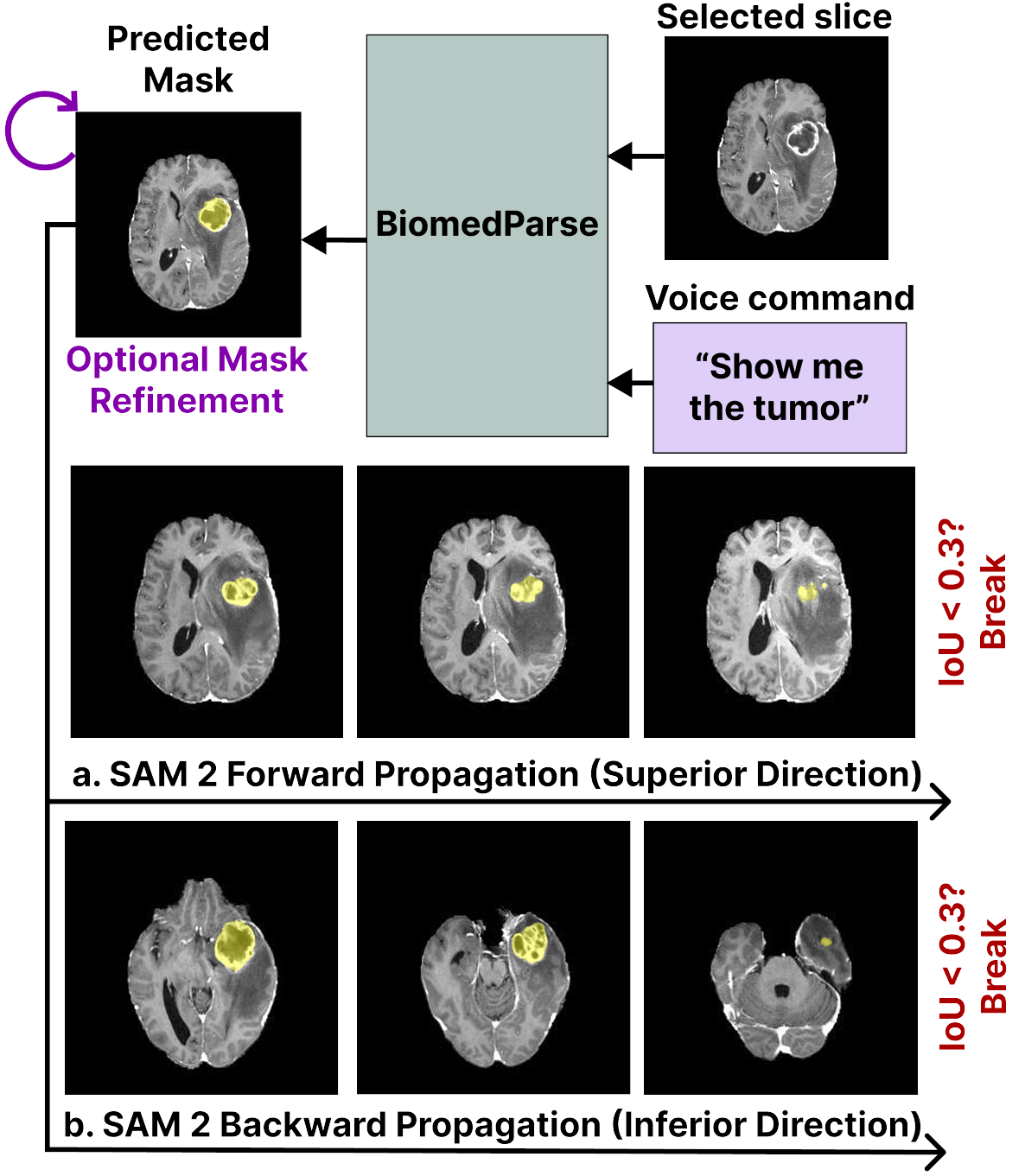}
    \caption{SAMIRA's segmentation module for mask prediction, refinement, and propagation across frames. After a voice command initiates initial tumor segmentation via BiomedParse, the user may optionally refine the mask through point prompts. The mask is then propagated slice-wise using SAM2, first superiorly (a) and then inferiorly (b), with propagation automatically terminating when inter-slice Intersection-over-Union (IoU) falls below 0.3 to prevent segmentation drift.}
    \label{fig:segmentation_module}
\end{figure}

We proposed a novel deep learning-based, speech-initiated interactive segmentation method for 3D medical images as illustrated in Figure \ref{fig:segmentation_module}. This method forms a key function of SAMIRA and relies on two complementary foundation models. The first model, \textit{BiomedParse}, generates the initial segmentation mask in response to user-issued voice commands. Spoken prompts are first transcribed using the Azure Speech AI service, producing a text prompt that is passed to BiomedParse along with the image slice. Upon receiving the textual prompt and corresponding image slice, BiomedParse  produces a binary mask of the target structure described in the user's voice command. The second model, SAM2, is used for interactive refinement and multi-slice propagation. Users can iteratively correct the mask using positive and/or negative point prompts, which are passed to SAM2 for real-time refinement. Once satisfied, the refined mask is propagated bi-directionally through the volume using a modified version of SAM2's memory mechanism with a novel propagation termination criterion: if the inter-slice Intersection-over-Union (IoU) between the current mask ($\text{Mask}_{t}$) and previous mask ($\text{Mask}_{t-1}$) fell below 0.3, propagation is halted in that direction, under the assumption that mask changes should be relatively smooth between neighboring image slices. Inter-slice IoU measures how much two sequentially predicted binary masks overlap, defined as the ratio between the area of their intersection and the area of their union:
\[
\text{IoU} = \frac{|\text{Mask}_{t} \cap \text{Mask}_{t-1}|}{|\text{Mask}_{t} \cup \text{Mask}_{t-1}|}
\]

While point-prompt-based revision with SAM2 helps ensure the accuracy of the segmentation, it is desirable to keep the needs at minimum for the efficiency of the workflow. Thus, to gauge the baseline performance of our proposed segmentation method in the absence of prompt-based revision, we conducted a standalone ablation study using the Pretreat-MetsToBrain-Masks \cite{ramakrishnan2024large} and LiTS \cite{bilic2023liver} datasets, by comparing the accuracy of a ``BiomedParse seeding + SAM2 propagation with IoU-based early stopping" pipeline versus a ``BiomedParse seeding + original SAM2 propagation" one. To accomplish this, we randomly selected 40 volumes (20 brain tumor MRI scans and 20 liver tumor CT scans) from the datasets, excluding cases used in the full workflow study. For each volume, a slice with visible tumor was manually selected and submitted to the inference server along with a natural language prompt —``show me the brain tumor" for brain cases and ``show me the liver tumor" for liver cases. Here, BiomedParse generates an initial tumor mask for the selected slice, which was then propagated bi-directionally (in the superior and inferior directions) using SAM2. Using the paired sample Wilcoxon signed-rank test, results indicate that the break condition significantly improved accuracy for liver tumor CT scans ($p = 0.0024 < 0.05$), while having a smaller but still significant effect for brain tumor MRIs ($p = 0.0039 < 0.05$), as summarized in Table~\ref{tab:iou-performance}. The stronger effect observed in CT scans may be attributed to the higher levels of noise and lower soft-tissue contrast, which can cause SAM2, originally trained on natural images, to mistake noise for anatomical structures. The break condition prevents propagation of spurious segmentations across slices, which is particularly helpful in noisier CTs. Compared to recent SAM2-based methods for 3D medical segmentation \cite{shen2025interactive3dmedicalimage}, our approach introduces both a language-based initialization and a propagation stopping rule, enhancing accuracy and reducing user burden. These findings support the inclusion of the IoU threshold as an essential mechanism for robust slice-wise mask propagation in clinical datasets.

\begin{table}[h!]
\centering
\caption{3D Dice scores (mean$\pm$std) for automatic SAM2 propagation with and without the IoU break condition.}
{\small 
\renewcommand{\arraystretch}{1.3} 
\begin{tabular}{|l|l|c|c|}
\hline
\textbf{Mod.} & \textbf{Target} & \makecell{\textbf{IoU}\\\textbf{Break} \rule{0pt}{2.8ex}} & \textbf{3D Dice} \\
\hline
\multirow{2}{*}{MRI} & \multirow{2}{*}{Brain Tumor} & \textbf{True}  & \textbf{87.41 $\pm$ 10.77}$^\dagger$ \\
                     &                              & False & 87.28 $\pm$ 10.73 \\
\hline
\multirow{2}{*}{CT}  & \multirow{2}{*}{Liver Tumor} & \textbf{True}  & \textbf{73.31 $\pm$ 13.04}$^\dagger$ \\
                     &                              & False & 68.94 $\pm$ 16.30 \\
\hline
\end{tabular}
}
\vspace{1mm}
\footnotesize{$^\dagger$Statistically significant improvement compared to no IoU break condition}
\label{tab:iou-performance}
\end{table}

\subsubsection{\textbf{RAG-based guidance system}}

In addition to interactive segmentation, to provide multi-modal, context-aware guidance during segmentation, SAMIRA employs a RAG framework that integrates image similarity search with generative language modeling. At its core is FAISS~\cite{douze2024faiss}, a library developed by Meta for fast approximate nearest-neighbor search on high-dimensional vectors. FAISS enables real-time retrieval of anatomically similar tumor and non-tumor slices from large medical image datasets, based on high-dimensional vectors from the high-level feature maps output by the Res5 layer of BiomedParse's image encoder. These retrieved examples are used in two ways: first, to provide general contextual grounding using representative healthy and pathological slices (RAG Request 1) and second, to generate query-specific guidance based on the user’s spoken prompt and the current image slice (RAG Request 2).

\begin{figure*}[t]
    \centering
    \includegraphics[width=\textwidth]{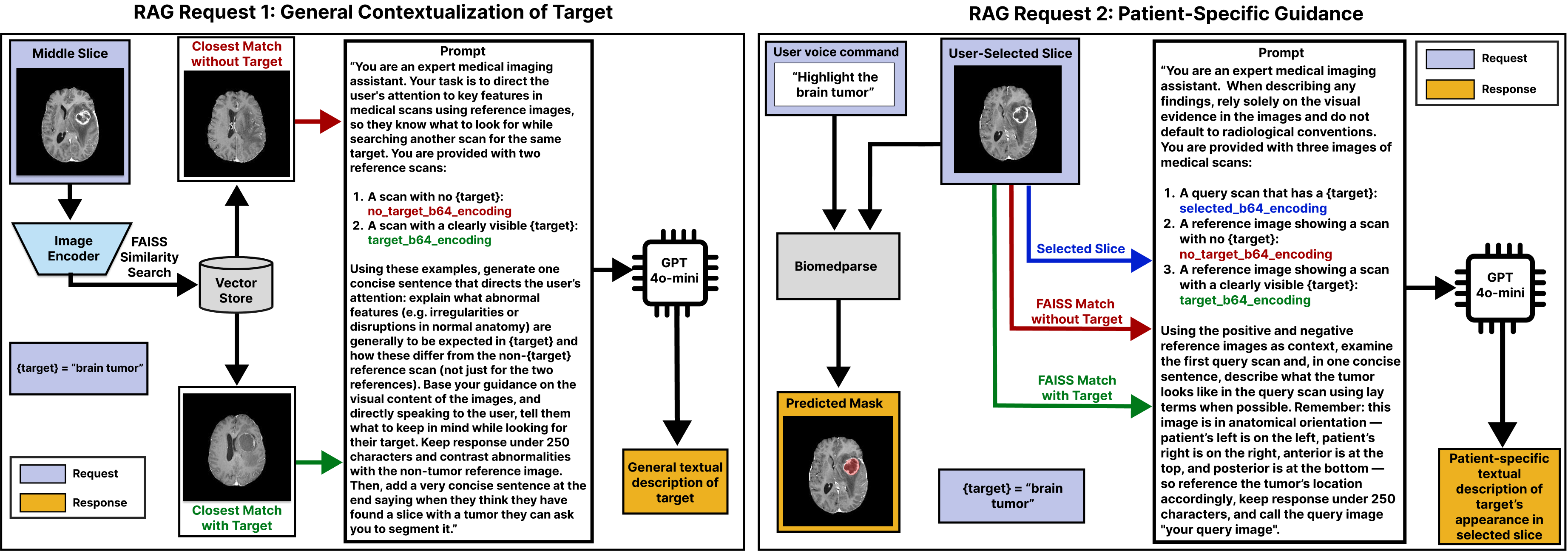}
    \caption{Retrieval-Augmented Generation (RAG) pipelines for multimodal guidance. (Left) To support initial understanding, the system retrieves two anatomically similar reference slices—one with and one without the target pathology—and uses them to generate a general description of the abnormality. (Right) After the user selects a slice and issues a voice command, the system compares visual features of the patient’s scan to healthy and pathological reference images. Guided by shared and differing features, the agent describes what the abnormality likely looks like in the selected slice.}
    \label{fig:RAG}
\end{figure*}

\noindent
\textbf{Knowledge Database Construction:} We constructed two FAISS databases, one for brain MRIs and one for liver CTs, which contain a total of 30,845 brain slices (6,766 tumor, 24,079 without tumor) and 57,193 liver slices (6,982 tumor, 50,210 without tumor), drawn from 199 and 127 patients, respectively. Embedding vectors were computed using BiomedParse, normalized, and stored for efficient similarity search. These databases can easily be expanded in the future with a larger variety of targets and cases.

\vspace{0.2cm}
\noindent
\textbf{RAG Request 1. Initial Contextualization:} When a new scan of interest is loaded in the system, SAMIRA uses the encoding vector of the middle slice to retrieve two visually similar reference images from the knowledge database: one with and one without the target pathology (i.e., tumor). These contrastive examples help ground the task as the non-tumor slice illustrates normal anatomy, while the tumor case highlights typical features of the target pathology. Then, both are passed to GPT-4o-mini, chosen for its fast inference time, which generates a general explanation of what to look for during visual exploration of the target. Figure~\ref{fig:RAG} illustrates this process, showing how supplying reference images with a carefully engineered prompt can support contextual grounding to guide the user.

\vspace{0.2cm}
\noindent
\textbf{RAG Request 2. Query-Specific Interpretation:}
While general contextualization offers users an overview of what the pathology typically looks like, it does not account for anatomical variation in individual cases being investigated. Therefore, query-specific interpretation is used to provide more personalized, patient-specific guidance. When the user issues a voice command to segment a structure (e.g., ``highlight the brain tumor"), the selected image slice is used to query the knowledge database, retrieving two visually similar reference slices: one with the target pathology (positive example) and one without (negative example). These references, \textit{along with the user’s query slice}, are passed to GPT-4o-mini in another structured multi-modal prompt, illustrated in Figure~\ref{fig:RAG}. By explicitly contrasting the \textit{query} with both healthy and pathological references, the agent can generate targeted explanations grounded in the anatomy of the target image under study.

\subsection{User Study and Evaluation Metrics}

We conducted two separate user studies for the developed system. While the first study aimed to identify the optimal user interaction paradigm for SAM2-based segmentation refinement, the second study assessed the full workflow of the proposed system. All user studies were conducted under institutional ethics approval and after all participants provided informed consent.

\subsubsection{\textbf{User-Interaction Paradigm Evaluation}}

We recruited 15 participants (age = $26.5 \pm 2.68$ years; 6 female \& 9 male) to evaluate the three interaction paradigms (\textit{Controller}, \textit{Head Pointing}, and \textit{Eye Tracking}). Participants rated their familiarity with VR, human anatomy, medical imaging modalities, and medical image segmentation on a 1–5 scale (1 = unfamiliar, 5 = familiar). On average, they reported familiarity with VR ($4.2 \pm 0.94$), and between neutral and somewhat familiarity with anatomy ($3.47 \pm 1.24$), imaging modalities ($3.53 \pm 1.30$), and medical image segmentation ($3.73 \pm 1.27$). Prior to the study, participants completed a guided hands-on tutorial in the VR environment, which included practice with segmentation mask refinement on a lung CT scan from the LCTSC dataset~\cite{LCTSC} and menu interaction using each of the three interaction paradigms. The tutorial session also served to calibrate the eye-tracking system. Following this, users began the interaction test, correcting the erroneous BRATS brain tumor MRI slices with each interaction paradigm as described in Section IV.A. For each interaction paradigm trial, task completion time, segmentation accuracy (with 3D Dice score), perceived task load, point prompts placed, and point prompts erased were recorded.

To assess perceived task load, we used the \textit{NASA Task Load Index (NASA-TLX)}~\cite{hart1988development}, a validated six-item instrument measuring mental demand, physical demand, temporal demand, effort, frustration, and perceived performance. Each item was rated in the range of 1-21, which was scaled to 0–100 for analysis, and finally all items were averaged to produce a total task load score. To compare overall performance across interaction paradigms, we computed a composite interaction score for each of the 45 user-paradigm trials (15 participants × 3 interaction paradigms). This score quantifies trade-offs between segmentation accuracy, task load, and completion time. Z-scores for accuracy, NASA-TLX, and completion time were calculated across all trials to ensure standardized comparison. For each trial $i$, the composite score was calculated as:

\[
\text{Composite}_i = z_\text{accuracy} - z_\text{NASA} - z_\text{time}
\]
\noindent
This score treats accuracy as beneficial and both task load and time as costs, giving equal weight to each. Mean and standard deviation of composite scores were then computed per interaction paradigm to summarize performance. Finally, for additional self-reported confirmation, we asked users to rank their preferred interaction paradigm.

\subsubsection{\textbf{Full Workflow Study of SAMIRA}}

To evaluate the complete AI-assisted segmentation system, including conversational interaction, mask refinement, and 3D visualization, we recruited 19 participants (age = $26.8 \pm 3.63$ years; 8 female \&  11 male). On average, they again reported familiarity with VR ($4.21 \pm 1.18$) and between neutral to somewhat familiarity with anatomy ($3.42 \pm 1.22$), imaging modalities ($3.79 \pm 1.18$), and medical image segmentation ($3.74 \pm 1.37$). After a brief tutorial, where participants practiced the full workflow by segmenting a lung CT scan from the LCTSC dataset \cite{LCTSC}, participants completed the two segmentation tasks: brain tumor in MRI and  liver tumor in CT. The order of these was randomized across participants to minimize order effects. Outcome measures included segmentation accuracy (3D Dice scores before and after segmentation refinement), task completion time, and user experience metrics. Task load was assessed using NASA-TLX, similarly to the interaction paradigm evaluation. To assess the overall usability, we also administered the \textit{System Usability Scale (SUS)}~\cite{brooke1996sus}, a widely used 10-item questionnaire that yields a total score from 0 to 100. Scores above 68 are considered to indicate good usability~\cite{brooke2013sus}. Additionally, we developed a custom 9-item questionnaire to evaluate users' perceptions of AI agent guidance, reference images, 3D visualization, and user confidence in the workflow. The full list of questions are provided in Figure~\ref{fig:custom_questionnaire}. Items used a 5-point Likert scale from 1 (strongly disagree) to 5 (strongly agree). Finally, we included an open-ended section, where users could freely mention what they liked and/or disliked about the system, and further elaborate their semi-quantitative evaluations.

\subsubsection{\textbf{Statistical Analysis}}

For the interaction paradigm study, differences in NASA-TLX scores, completion times, and Dice scores across three user-interaction paradigms were evaluated using Kruskal–Wallis tests. For the full workflow study, we compared the 3D Dice scores before and after user refinement using Wilcoxon rank-sum tests. SUS scores were tested against the usability benchmark of 68 using a one-sample $t$-test. Custom user questionnaire responses were tested against the neutral value of 3 using Wilcoxon signed rank tests. For all thee statistical tests, a statistical significance was confirmed with \( p < 0.05 \).

\section{Results}\label{sec5}

\subsection{Interaction Paradigm Evaluation}


\begin{table*}[ht]

\centering
\caption{Comparison of interaction paradigms for segmentation refinement.
         Values are mean\,$\pm$\,standard deviation.
         The best score is in bold.  NASA-TLX is out of 100.}
\label{tab:interaction-paradigms}

\renewcommand{\arraystretch}{1.15}   

\begin{tabular*}{\textwidth}{@{\extracolsep{\fill}}lcccc@{}}
\toprule
\textbf{Interaction Mode} & \textbf{3D Dice (\%)} & \textbf{Time (s)} & \textbf{NASA-TLX} & \textbf{Composite Score} \\

\midrule
Controller    & \textbf{99.25 $\pm$ 0.25} & \textbf{220.3 $\pm$ 79.3} & 18.8 $\pm$ 14.5 & \textbf{0.51 $\pm$ 1.91}\\
Head Pointing & 99.21 $\pm$ 0.30          & 248.8 $\pm$ 78.5          & \textbf{16.8 $\pm$ 13.9} & 0.20 $\pm$ 1.56\\
Eye Tracking  & 99.13 $\pm$ 0.46          & 251.1 $\pm$ 78.7          & 26.6 $\pm$ 15.4 & –0.71 $\pm$ 1.77\\
\bottomrule
\end{tabular*}
\end{table*}

\subsubsection{\textbf{Accuracy, Completion Time, and NASA-TLX}}

\begin{figure}[h!]
  \centering
    \includegraphics[width=1\linewidth]{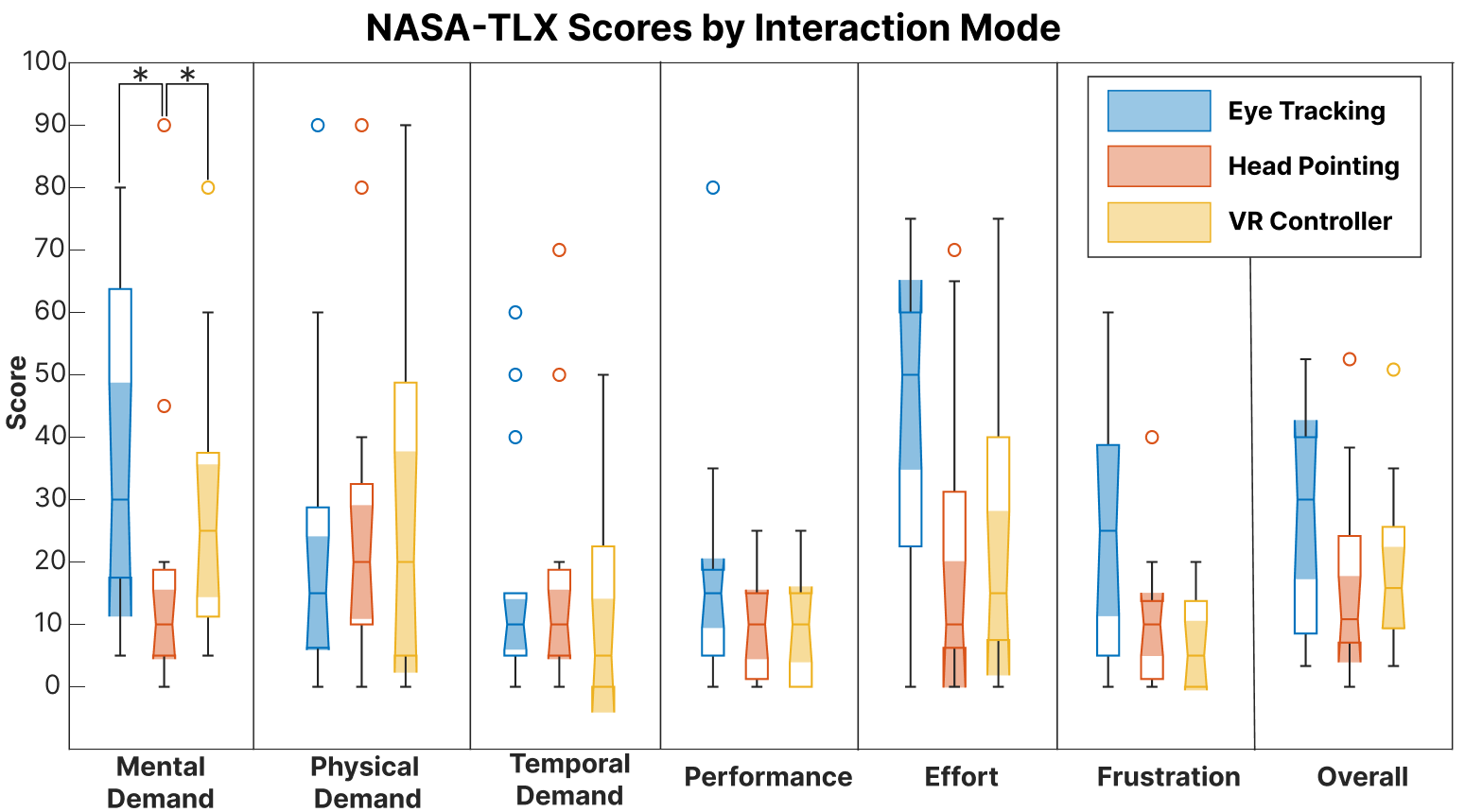}
  \caption{Boxplots of NASA Task Load Index (TLX) for different interaction paradigms, including Controller, Head Pointing, and Eye tracking. Significant pair-wise differences are marked with ``*".}
  \label{fig:nasa_interactions}
\end{figure}

 The metrics for evaluating the three interaction paradigms are shown in Table ~\ref{tab:interaction-paradigms}, with non-significant differences in overall scores between groups ($p>0.05$). For \textbf{3D Dice (segmentation accuracy)}, all paradigms yielded highly accurate refined masks, significantly above the starting 3D Dice score of 0.91 ($p<0.05$). The \textit{Controller} paradigm had the highest accuracy score with the least variability ($99.25 \pm 0.25$\%), followed by \textit{Head Pointing} ($99.21 \pm 0.30$\%), then \textit{Eye Tracking} ($99.13 \pm 0.46$\%). For \textbf{completion time}, the ranking remained the same: \textit{Controller} was the fastest ($220.3 \pm 79.3s$) followed by \textit{Head Pointing} ($248.8 \pm 78.5s$) and \textit{Eye Tracking} ($251.1 \pm 78.7s$). The overall \textbf{NASA-TLX} scores (out of 100) slightly favored \textit{Head Pointing}, which had the lowest overall task load ($16.8 \pm 13.9$), followed by \textit{Controller} ($18.8 \pm 14.5$) and \textit{Eye Tracking} ($26.6 \pm 15.4$). While overall task load did not demonstrate significant differences, one sub-item showed a significant effect: mental demand was significantly lower for \textit{Head Pointing} ($16.7 \pm 22.5$) compared to \textit{Controller} ($30.0 \pm 23.9$, $p=0.0123$) and \textit{Eye Tracking} ($38.7 \pm 25.5$, $p=0.0329$), as illustrated in Figure~\ref{fig:nasa_interactions}. No other NASA-TLX sub-items' differences reached statistical significance.

\subsubsection{\textbf{Composite Scores}} The resulting mean composite scores that promote a balance between performance, speed, and user effort were $0.51 \pm 1.91$ for \textit{Controller}, $0.20 \pm 1.56$ for \textit{Head Pointing}, and $-0.71 \pm 1.77$ for \textit{Eye Tracking}. Among the paradigms, \textit{Controller} achieved the highest mean composite score, driven by a favorable combination of high segmentation accuracy ($99.25$\%), low task load (NASA-TLX = $18.8$), and the fastest completion time ($220.3$ s). \textit{Head Pointing} followed closely, while \textit{Eye Tracking} lagged behind due to relatively higher task load and slower performance.

\subsubsection{\textbf{Point Prompt Efficiency}}
To complement the accuracy, task completion time, and NASA-TLX, we also compared the three interaction paradigms in terms of the ratio between the number of confirmed point prompts for segmentation refinement and the cleared points (i.e., points-per-clear efficiency, the higher the better). This ratio reveals the robustness of these paradigms with the joint consideration of their inherent precision, efficiency, and task load. Overall, \textit{Controller} yielded the highest points-per-clear efficiency ($24.33 \pm 24.91$), followed by \textit{Head Pointing} ($16.08 \pm 6.17$) and \textit{Eye tracking} ($14.67 \pm 8.72$), suggesting that the \textit{Controller} condition was more robust for point prompt placement (less corrections required) and eye tracking was more prone to errors.

\subsubsection{\textbf{User preferences}} Finally, when asking the participants to rank their preferred interaction paradigms, 7 of 15 participants selected \textit{Controller}, 6 selected \textit{Head Pointing}, and only 2 selected \textit{Eye tracking} as their top choices. These results, in addition to the close composite scores, suggest that both \textit{Controller} and \textit{Head Pointing} are viable and well-received paradigms, while \textit{Eye tracking} is comparatively less favored. The close ranking between \textit{Controller} and \textit{Head Pointing} indicates that either approach could be suitable for deployment in the full workflow. However, to reduce variability in downstream evaluation, we selected \textit{Controller} as the primary interaction paradigm for the subsequent full workflow study due to its marginally superior results.

\subsection{Full Segmentation Workflow with SAMIRA}

\subsubsection{\textbf{Segmentation accuracy}} 
In terms of segmentation accuracy (3D Dice), the workflow yielded high scores. For the \textbf{brain MRI data}, refined masks (94.92~$\pm$~0.52\%) showed significantly higher accuracy than the user's propagated unrefined masks (90.53~$\pm$~10.51\%, $p < 0.0001$). For the \textbf{liver CT data}, Dice scores were comparable between unrefined (95.47~$\pm$~0.39) and refined (95.46~$\pm$~0.40) masks, with no significant difference, ultimately indicating high starting accuracies are preserved (Table \ref{tab:dice_scores}). This high starting accuracy is likely attributed to the liver tumor case being larger and more spherical shaped than the brain tumor case that appears as a mass centrally, but splits into multiple lobes in the superior and inferior regions. The lower difficulty of the liver tumor case is further reflected in the shorter completion time (liver: $279.6 \pm 109.4$s vs. brain: $361.7 \pm 144.3$ s) and fewer point prompts placed (liver: 8.8 $\pm$ 6.5 vs. brain: 27.7 $\pm$ 19.9 ). The brain tumor case likely demanded more user input due to its aforementioned irregular shape.

\subsubsection{\textbf{Semi-quantitative questionnaire results}} 
Participants rated the overall workflow with SAMIRA as highly usable. The \textbf{System Usability Scale (SUS)} score was $90.00 \pm 8.98$, which is significantly higher than the benchmark of 68 ($p < 0.001$) and corresponds to an `A' usability score \cite{brooke2013sus}. Meanwhile, the \textbf{NASA-TLX} scores (out of 100) indicate low to moderate task loads across sub-items. Descriptive statistics are as follows: Mental Demand (31.84~$\pm$~28.10), Physical Demand (12.11~$\pm$~14.27), Temporal Demand (13.68~$\pm$~18.25), Performance (20.00~$\pm$~24.72), Effort (25.26~$\pm$~15.50), Frustration (7.11~$\pm$~8.22), and Overall task load (18.33~$\pm$~11.93).

\begin{table}[h!]
\centering
\caption{3D Dice scores before and after refinement. Asterisks denote statistically significant changes.}
\begin{tabularx}{\linewidth}{|l|X|X|X|}
\hline
\textbf{Tumor} & \textbf{Unrefined} & \textbf{Refined} & \textbf{p-value} \\
\hline
Brain & $90.53\pm10.51$ & $94.92 \pm 0.52$ & $1.19 \times 10^{-5}$* \\
Liver & $95.47 \pm 0.39$ & $95.46 \pm 0.40$ & 0.3968 \\
\hline
\end{tabularx}
\label{tab:dice_scores}
\end{table}

\begin{figure}[h!]
  \centering
    \includegraphics[width=1\linewidth]{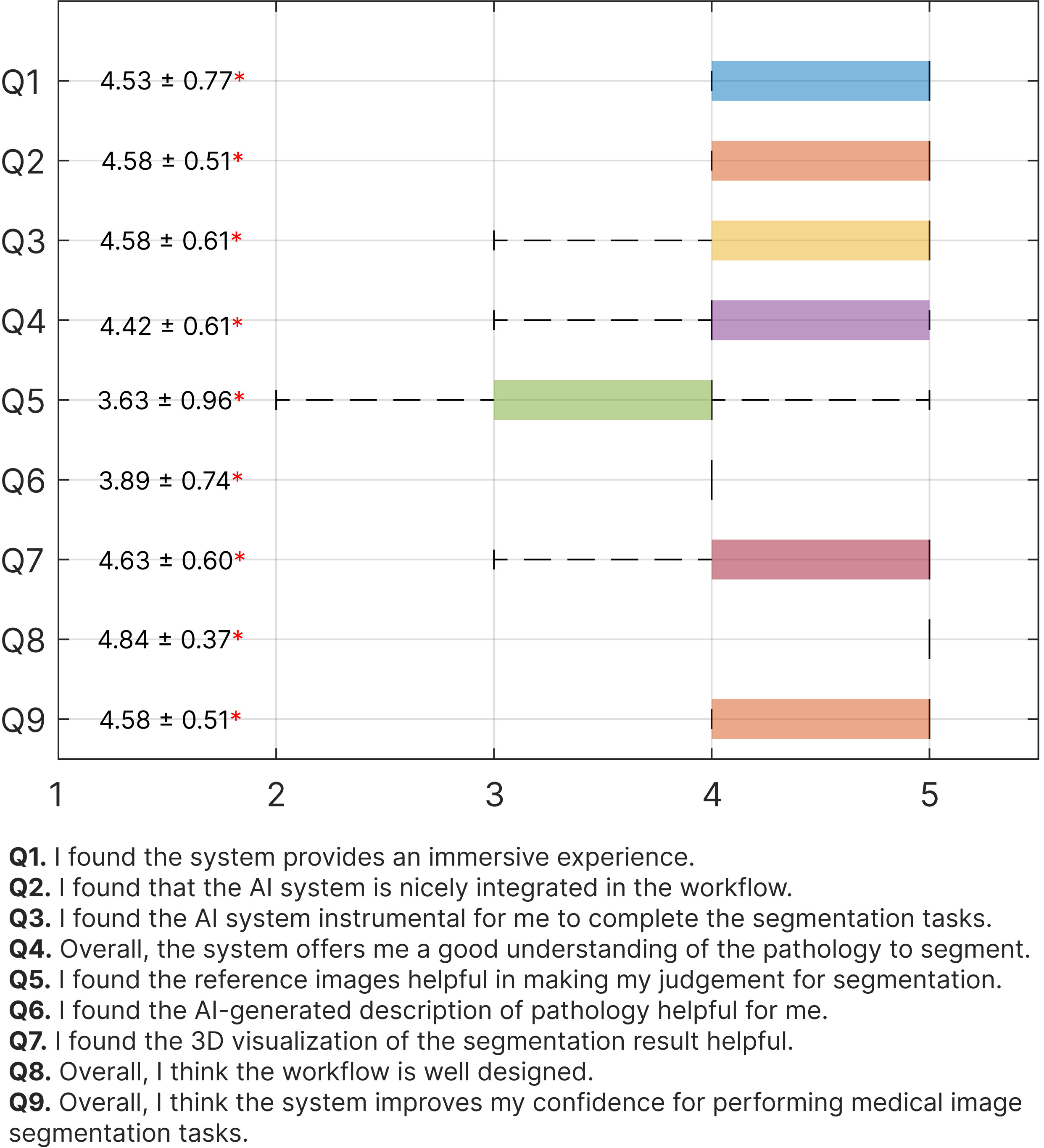}
  \caption{Boxplots of custom user experience questionnaire results (mean ± std on left, values significantly above 3 with red asterisk) for the full workflow with SAMIRA.} 
  \label{fig:custom_questionnaire}
\end{figure}

Finally, all responses on the \textbf{custom questionnaire} were significantly above the neutral midpoint of 3 ($p < 0.05$), indicating favorable perceptions of the system's AI agent integration, its ability to support pathology learning, the helpfulness of reference images, and the usefulness of 3D visualization (see Figure~\ref{fig:custom_questionnaire}).

\subsubsection{\textbf{Qualitative feedback results}}

Participants' written feedback, where they were asked to describe what they liked or disliked about the system, further supports the system's perceived usability and educational value. \textit{16 out of 19 participants} described the system as ``easy to use,” ``intuitive,” or ``clear,” highlighting its low learning curve and smooth integration with the VR environment. One participant noted, ``After you learn it, it is easy to use, fast, and interactive,” while another stated, ``The system is very user friendly and well integrated with the HTC Vive controls.” These perceptions align with the high overall scores on SUS Question 3 (``I thought the system was easy to use," 4.58 ± 0.51) and Question 5 (``I found the various functions in this system were well-integrated," 4.68 ± 0.48). AI voice interaction was also frequently praised. \textit{10 participants} explicitly mentioned that the voice command feature improved usability, with one stating, ``The voice commands are well integrated and the propagation is very helpful for identifying the whole tumor.” Another remarked, ``The AI was well integrated and didn’t feel intrusive—more like a helpful assistant." The system was also perceived as supportive of learning and decision-making. \textit{7 participants} reported increased confidence during segmentation tasks, attributing this to guidance from the AI agent and the real-time visualization tools. For example, one wrote, “It was so useful and made me confident to do the task”, echoing high SUS Question 9 results (“I felt very confident using this system,” 4.74 ± 0.45).

While overall impressions were positive, some participants suggested providing additional medical context, and a few expressed awkwardness with voice inputs. Nonetheless, the qualitative feedback reinforces the questionnaire findings and confirms that SAMIRA’s design successfully balances guidance, autonomy, and interpretability in a VR environment.

\section{Discussion}\label{sec6}

The findings of our experiment provided partial support for our hypothesis that \textit{Head Pointing} would provide the optimal trade-off between segmentation accuracy, efficiency, and task load. While \textit{Head Pointing} did show the lowest mental demand, \textit{Controller}-based input achieved slightly better overall performance, as reflected in its higher mean accuracy and efficiency metrics and its highest average composite score. Furthermore, all paradigms yielded excellent Dice scores following segmentation correction, but \textit{Controller} and \textit{Head Pointing} outperformed \textit{Eye Tracking} in terms of accuracy, task load, and completion time, echoing the findings of Xu et al.'s evaluation of text-selection techniques~\cite{xu2022evaluation}, where head-pointing and controller performance were close. Despite composite scores favoring controller-based pointing overall, \textit{Head Pointing} emerged as a lower mental effort alternative—especially for applications where users may need to work with just one hand or seek a cognitively lighter interaction. \textit{Eye Tracking}, while promising in theory, remains less favored for segmentation refinement tasks, where precision and visual stability are critical, despite the damping function (Equation \ref{eq:smoothing}) we employed to improve precision and user-experience. In future deployments, the system can allow users to select their preferred interaction paradigm, offering flexibility. This flexibility would be warranted, since all three paradigms were chosen as favorites across the different users.

Overall, our findings suggest that AI-assisted segmentation in VR is not only technically viable, but also educationally and ergonomically impactful. Across studies, users were able to generate high-quality segmentations without domain expertise and the system demonstrated that it promotes understanding, confidence, and informed interaction. In the full workflow, participants rated the system highly on both usability and interpretability of the task. As seen in Figure~\ref{fig:custom_questionnaire}, users strongly agreed that the AI agent helped them complete tasks (Q3: 4.58±0.61), supported understanding of the pathology (Q4: 4.42±0.61), and improved their confidence in performing segmentation tasks (Q9: 4.58±0.51). The RAG mechanism played a key role here. By comparing the queried radiological slice to both healthy and pathological references, users received contextualized, anatomy-specific guidance that was grounded in real cases. This was especially reflected in high agreement with Q8, where workflow design was rated highest (4.84±0.37). Interestingly, the AI-generated textual explanations (Q6) and reference images (Q5) received slightly lower scores (3.89±0.74 and 3.63±0.96, respectively), possibly reflecting uncertainty and confusion in some participants, who in general did not have strong familiarity with human anatomy ($3.42 \pm 1.22$, 1 = unfamiliar, 5 = familiar). To improve comfort and presence during voice interaction, future versions of the system could feature a visual avatar for the assistant, helping to reduce the slight awkwardness some users felt when speaking to a disembodied voice.

Furthermore, it was evident in the segmentation results that users demonstrated a clear understanding of when to intervene and when to trust the AI system. For example, with the simpler liver tumor CT case, users made minimal edits, and the refined masks were similar to unrefined ones. Yet importantly, performance did not degrade after user interaction, indicating that users did not over-correct or introduce noise. This suggests a healthy level of trust and restraint, and a true understanding of segmentation quality based on visual features and reference images. In contrast, the more challenging brain tumor MRI case, which included branching outer boundaries, showed a significant accuracy improvement after refinement. The 3D Dice score rose from $90.53\pm10.51$\% to $94.92\pm0.52$\% post-interaction ($p < 0.001$), demonstrating that users could identify areas needing correction and effectively apply point-based refinements. This reinforces that users not only understand segmentation correctness, but can also meaningfully enhance AI outputs in cases where human expertise is still necessary.

Together, these results position our system not only as a segmentation tool, but as a supportive assistant capable of accelerating workflows, teaching radiological features of pathology, and fostering trust with clinical AI. The positive responses to confidence and task understanding, combined with high segmentation accuracy, suggest that such a tool has potential in both clinical workflows and medical education. Future work may explore expanding the agent to work with more radiological concepts, or adaptive RAG responses based on user skill level (i.e. beginners versus experienced radiologists).

\section{Conclusion}\label{sec7}

We introduced a novel VR system for interactive medical image segmentation that integrates foundation models with attention-switching interaction and a supportive conversational AI agent, SAMIRA. At the core of our method is a novel segmentation algorithm that combines BiomedParse’s language-driven detection with a medical-image-adapted SAM2 model. To adapt SAM2 for clinical imaging, we introduced a novel IoU-based stopping criterion in its memory mechanism to prevent drift across noisy or low-contrast slices. Our findings show that this criterion can significantly improve segmentation quality, and that by using it with SAMIRA, users can efficiently achieve high segmentation accuracy with minimal effort across all interaction paradigms. Of these interaction paradigms, \textit{Controller} pointing offers the best overall balance of accuracy, speed, and task load, closely followed by head-pointing. More importantly, users demonstrated a clear ability to interpret and refine AI outputs based on its generated guidance, engaging critically with reference images, contextual explanations, and 3D visualization.

Importantly, this system is generalizable to any 3D medical image aligned with BiomedParse’s training scope and can be expanded by enriching its RAG knowledge database. Its modular design, high usability, and adaptability to various interaction styles position it as a powerful tool for clinical workflows, but also offers insights for future HCI research in intelligent, immersive medical systems with clinical AI agents.

\backmatter

\bmhead{Acknowledgements}

We acknowledge the support of the Natural Sciences and Engineering Research Council of Canada (596537, RGPIN/05100-2022), the Fonds de recherche du Québec \textbf{–} Nature et technologies (B1X-348625), and the Fonds de recherche du Québec \textbf{–} Santé Junior 1 Research Scholar program. We thank Taha Koleilat for the insightful discussion on medical foundation models.

\section*{Declarations}



\noindent \textbf{Conflict of interest} The authors declare no conflict of interest. \\

\noindent \textbf{Data availability} The data that support the findings of this study are available from the corresponding author upon reasonable request. 

\bibliography{sn-bibliography}

\end{document}